\documentclass[journal]{IEEEtran}
\usepackage{amsmath,amsfonts}
\usepackage{amssymb} 
\usepackage{algorithmic}
\usepackage{algorithm}

\usepackage{array}
\usepackage[caption=false,font=normalsize,labelfont=sf,textfont=sf]{subfig}
\usepackage{textcomp}
\usepackage{stfloats}
\usepackage{url}
\usepackage{verbatim}
\usepackage{graphicx}
\usepackage{cite}
\usepackage{amsthm}
\theoremstyle{plain}

\usepackage{mathtools}  

\usepackage{booktabs}
\usepackage{siunitx}
\allowdisplaybreaks
\usepackage{booktabs}
\usepackage{hhline}
\usepackage[table]{xcolor}
\usepackage{diagbox}
\usepackage{makecell}
\usepackage{microtype} 
\newcounter{example}[subsection]
\renewcommand{\theexample}{\Roman{example}}
\newcommand{\example}{\refstepcounter{example}\textit{\textbf{Example \theexample:} }}

\usepackage{amsmath,amssymb,mathtools}

\definecolor{darkgreen}{rgb}{0.0, 0.673, 0.51}
\definecolor{darkcyan}{rgb}{0.0, 0.35, 0.55}

\usepackage[colorlinks=true, linkcolor=black, citecolor=black, urlcolor=black, pdfborder={0 0 0}]{hyperref}  

\newtheorem{remark}{Remark}

\newcommand{\eqa}{\mathrel{\stackrel{\scriptstyle\text{(a)}}{=}}}
\newcommand{\eqb}{\mathrel{\stackrel{\scriptstyle\text{(b)}}{=}}}

\newcommand{\orcidicon}[1]{\href{https://orcid.org/#1}{\raisebox{0.4ex}{\includegraphics[height=1.6ex]{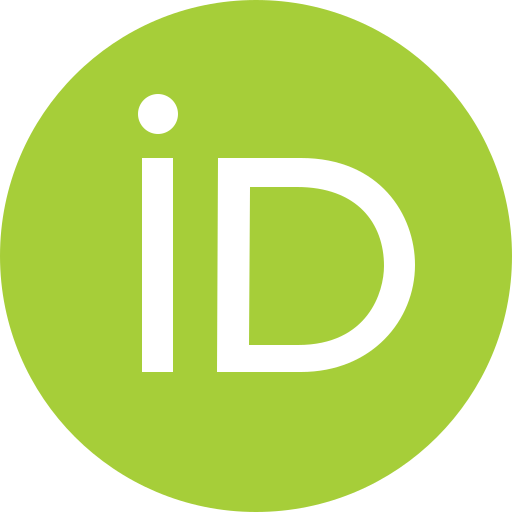}}}}

\usepackage{fancyhdr}

\fancypagestyle{firstpage}{%
  \fancyhf{} 
  
  \fancyhead[L]{\scriptsize 
 	This work has been submitted to the IEEE for possible publication. Copyright may be transferred without notice, after which this version may no longer be accessible. 
  }%
}

\begin{document}
\makeatletter
\typeout{IEEEtran text font size = \f@size pt}
\makeatother

\title{A Unified Analytical Nullspace-Based Least-Squares Design of the Farrow Structure} 
\author{Deijany Rodriguez Linares\orcidicon{0009-0004-1846-9496}, \textit{Graduate Student Member, IEEE}, and H\aa kan Johansson\orcidicon{0000-0001-6329-9132}, \textit{Senior Member, IEEE}%
\thanks{The authors are with the Department of Electrical Engineering, Linköping University, 58183 Linköping, Sweden. Email: \{deijany.rodriguez.linares, hakan.johansson\}@liu.se.}
\thanks{This work was funded by ELLIIT.}}

\maketitle
\thispagestyle{firstpage} 
\begin{abstract}
\sloppy
Farrow structures based on linear--phase FIR subfilters provide an efficient realization of variable fractional--delay (VFD) filters with reduced implementation complexity. While the all--linear--phase configuration admits a decoupled least--squares (LS) formulation with an analytical solution, this decoupling fails when branches of mixed types, linear--phase and general FIR, are required, as occurs when a group--delay constraint is imposed. This letter presents a unified LS design for Farrow structures via a nullspace parameterization of the per--branch symmetry constraints, yielding an analytical solution that accommodates arbitrary per--branch types. Numerical results demonstrate that the proposed framework satisfies group--delay constraints that the all linear--phase approach cannot meet, while substantially reducing the number of free parameters relative to the unconstrained general FIR baseline.
\end{abstract}

\section{Introduction}
\label{sec:intro}
\begin{figure}[tbp]
	\centering \includegraphics[scale=1.0]{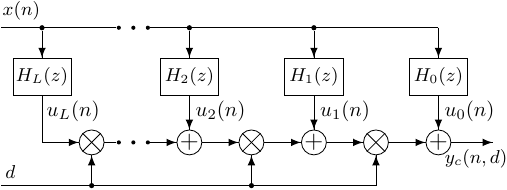}
	\caption{Farrow structure.} 
	\label{flo:farrow}
\end{figure}

\IEEEPARstart{V}ARIABLE fractional--delay (VFD) filters are widely used in applications such as sampling--rate conversion, timing synchronization, timing--mismatch calibration, and wideband beamforming~\cite{Chen2024, Liu2015, deiokshak2025, Shi_2023, Canese2023}. Polynomial-based implementations of VFD filters, known as Farrow structures~\cite{Farrow1988} (see Fig.~\ref{flo:farrow}), express the impulse response as a polynomial in the fractional--delay parameter, realized by parallel FIR subfilters. By requiring the subfilters to satisfy the linear-phase condition, i.e., to have a symmetric or antisymmetric impulse response~\cite{Hunter2009, Eghbali2013, Johansson2003}, the number of multiplications is significantly reduced while preserving interpolation accuracy, making these architectures attractive for hardware--efficient implementations.

Existing design methods for Farrow structures are typically based on a polynomial expansion of the desired fractional--delay response with respect to the delay parameter, where each branch corresponds to one expansion term. Classical approaches use Taylor or maximally flat approximations~\mbox{\cite{Laakso1996,Haolin2017,Haolin2019,Samadi2004}}, while later works apply minimax or sparsity--oriented optimizations to the expansion coefficients~\cite{Hunter2009,Johansson2003,Eghbali2013,Srivatsan2020}. 
Least--squares (LS) solutions have also been derived, for general unsymmetric branch filters~\cite{Tarczynski1997}, as well as linear--phase branch filters exhibiting coefficient symmetry~\cite{Deng2006} where the even/odd branch decoupling yields two separated closed--form solutions; 
however, this decoupling holds only when all branches have linear--phase symmetry, and breaks down when general unsymmetric branches are required as arises in latency-critical applications~\cite{Canese2023} where the filter group delay is explicitly constrained, a scenario that has been studied with minimax methods~\cite{Eghbali2011} but for which no closed--form LS solution has been reported. Similarly, the approach of~\cite{Tarczynski1997}, while providing a closed-form solution for all-unsymmetric designs, does not extend to mixed branch types.

The method proposed in this letter addresses this limitation by adopting a nullspace parameterization that directly enforces the Farrow--structure constraints without assuming any particular branch symmetry type. This yields a global least--squares formulation whose closed--form solution coincides with the results of~\cite{Tarczynski1997} and~\cite{Deng2006} as special cases, up to negligible numerical errors, and naturally addresses general and mixed branch types without any reformulation.

Following this introduction, Section~\ref{sec:mod_farrow} formulates the Farrow structure design problem. Section~\ref{sec:LS} derives the proposed least-squares closed--form solution via a nullspace parameterization. Numerical results illustrating the performance of the proposal are presented in Section~\ref{sec:results}, whereas Section~\ref{sec:conclusion} concludes the paper.

\section{Farrow Structure Design Problem} \label{sec:mod_farrow}
We consider the approximation of an ideal VFD frequency response
\begin{align}
    H_{\mathrm{des}}(e^{j\omega }, d) = e^{-j\omega (d+D)}, 
    \label{eq:Hdes}
\end{align}
over a frequency band $\omega {\in} \Omega$ and delay range $d {\in}\mathcal{D}$, and where $D$ is the desired fixed delay. To approximate \eqref{eq:Hdes}, we adopt the Farrow structure, as depicted in \mbox{Fig. \ref{flo:farrow}}, composed of $L{+}1$ FIR subfilters, where $d$ is the fractional delay parameter. The resulting transfer function is
\begin{align}
    H(z,d)
    &{=} \sum_{k=0}^{L} d^k H_k(z),
    \label{eq:HFarrow}
\end{align}
where each $H_k(z) = \sum_{n=0}^{N_k} h_k[n]\,z^{-n}$ is a causal FIR filter, of order $N_k$ for branch~$k$, and the branch orders $N_k$ may differ across branches.

The design objective is to minimize the approximation error in the least-squares sense over the frequency--delay domain,
\begin{align}
\min_{\{N_k\}_{k=0}^{L}}
\int_{\Omega}\int_{\mathcal{D}}
\Big|\hspace{-1 pt}H_{\mathrm{des}}(e^{j\omega },d)
      {-} \sum_{k=0}^{L} \hspace{-2 pt}d^{k} H_k(e^{j\omega})\Big|^2 \,\mathrm{d}d\,\mathrm{d}\omega.
\label{eq:HFarrow_opt}
\end{align}

 To halve the number of design coefficients, one may impose linear-phase symmetry on all branches, which decouples the LS objective in~\eqref{eq:HFarrow_opt} into independent cosine and sine subsystems, each admitting a closed-form solution~\cite{Deng2006}. However, as mentioned in Section~\ref{sec:intro}, this decoupling breaks down when mixed branch types are required---as occurs when the group delay is constrained and some branches must be realized as general FIR subfilters---and no closed-form solution has been reported for this case.

\section{Proposed LS Design via Nullspace Constraint} \label{sec:LS}
The integral in \eqref{eq:HFarrow_opt} is approximated by sampling on a finite frequency--delay grid, $\omega_i$, $i=1,\dots,N_\omega$, and $d_\ell$, $\ell=1,\dots,N_d$. For a given set $\{N_k\}$, setting $H(e^{j\omega_i},d_\ell)=H_{\mathrm{des}}(e^{j\omega_i T},d_\ell)$ yields
\begin{align}
\sum_{k=0}^{L} d_\ell^{k} \sum_{n=0}^{N_k} h_k[n] e^{-j\omega_i n}
{=} e^{-j\omega_i (T d_\ell {+} D)}.
\label{eq:LS}
\end{align}

Stacking \eqref{eq:LS} for all $(i,\ell)$ results in the linear system $\mathbf{A}\boldsymbol{\theta}=\mathbf{b}$, whose least--squares solution is
\begin{align}
\boldsymbol{\theta}^\star
= \arg\min_{\boldsymbol{\theta}}\ \|\mathbf{A}\boldsymbol{\theta}-\mathbf{b}\|_2^2
= (\mathbf{A}^{\mathsf H}\mathbf{A})^{-1}\mathbf{A}^{\mathsf H}\mathbf{b},
\label{eq:unconstrained}
\end{align}
where, for $M{=}N_\omega N_d$ and $P{=}\sum_{k=0}^{L}(N_k{+}1)$, $\mathbf{A}$ is full column rank provided $M \geq P$ and the frequency--delay grid is nondegenerate.\footnote{\label{foo:grid}The grid $\{\omega_i\}_{i=1}^{N_\omega}\times\{d_\ell\}_{\ell=1}^{N_d}$ is nondegenerate if the $P$ columns of $\mathbf{A}$ are linearly independent; this holds in general for distinct $\omega_i$ and distinct $d_\ell$ whenever $M\geq P$ due to the exponentials in \eqref{eq:LS}.} The coefficient vector $\boldsymbol{\theta} \in \mathbb{R}^P$ is defined by
\begin{align}
      \boldsymbol{\theta}=
      \begin{bmatrix}
            \underbrace{h_0[0] \cdots h_0[N_0]}_{\mathbf{h}_0} \ \ \cdots \ \
            \underbrace{h_L[0] \cdots h_L[N_L]}_{\mathbf{h}_L}
      \end{bmatrix}^{\!\mathsf T}
      \in\mathbb{R}^{P},
\end{align}
and the entries of $\mathbf A\in\mathbb{C}^{M\times P}$, and $\mathbf b\in\mathbb{C}^M$ are given by
\begin{align}
      \mathbf{A}_{(i,\ell),\,(k,n)} &= d_\ell^{k}e^{-j\omega_i n},
\end{align}
\begin{align}
      \mathbf{b}_{(i,\ell)} &= e^{-j\omega_i (T d_\ell + D)}. \label{eq:b}
\end{align}
Here, each frequency--delay pair $(i,\ell)$ indexes one row in the least--squares system and each branch--tap pair $(k,n)$ one column.\footnote{That is, $(i,\ell) \triangleq (i-1)N_d+\ell$ and $(k,n) \triangleq \textstyle\sum_{{k'}=0}^{k-1}(N_{k'}{+}1)+n+1$, respectively.}

To incorporate per-branch constraints into \eqref{eq:unconstrained}, each branch $k$ is assigned a constraint matrix $\mathbf{C}_k \in \mathbb{R}^{r_k \times (N_k+1)}$, with $0 \leq r_k \leq \lfloor {N_k/2 }\rfloor{+}1$, such that $\mathbf{C}_k \mathbf{h}_k = \mathbf{0}$ encodes linear-phase symmetry. For Type-I (symmetric, even $N_k$) and Type-III (antisymmetric, even $N_k$) branches, the explicit forms are:
\begin{align}
\mathbf{C}^{{\mathrm{I}}}_k
&=
\begin{bmatrix}
1 & 0 & \cdots & 0 & 0&0 & \cdots & 0 & -1\\
0 & 1 & \cdots & 0 &0& 0 & \cdots & -1 & 0\\[4pt]
\vdots & \vdots & \ddots & \vdots & \vdots & \vdots & \ddots & \vdots &\vdots\\[4pt]
0 & 0 & \cdots & 1 &0& -1 & \cdots & 0 & 0 \\
\end{bmatrix}, \quad \text{and}
\\
\mathbf{C}^{{\mathrm{III}}}_k
&=
\begin{bmatrix}
1 & 0 & \cdots & 0 & 0& 0 & \cdots & 0 & 1\\
0 & 1 & \cdots & 0 & 0& 0 & \cdots & 1 & 0\\
\vdots & \vdots & \ddots  & \vdots & \vdots& \vdots &  \ddots      & \vdots & \vdots\\
0 & 0 & \cdots & 1 &0& 1 & \cdots & 0 & 0\\[4pt]
0 & 0 & \cdots & 0 &1& 0 & \cdots & 0 & 0
\end{bmatrix}.
\end{align}
The analogous forms for $\mathbf{C}^{{\mathrm{II}}}_k$ (Type-II: symmetric, odd $N_k$) and $\mathbf{C}^{{\mathrm{IV}}}_k$ (Type-IV: antisymmetric, odd $N_k$) are constructed similarly. In all cases, for a branch $k$ of order $N_k$, the entries of $\mathbf{C}_k$ are given by
\begin{align}
\mathbf{C}_k(m,n) &= \delta(n-m) + \sigma_k\,\delta\!\bigl(n-(N_k-m)\bigr), \label{eq:Ck_entries}\\
\mathbf{C}_k\!\left(\tfrac{N_k}{2},\,n\right) &= \delta\!\left(n-\tfrac{N_k}{2}\right),\quad\text{Type-III only ($N_k$ even),}
\end{align}
for $m=0,\dots,\lceil N_k/2\rceil-1$, $n=0,\dots,N_k$, where $\sigma_k{=}{-}1$ for symmetric branches (Types~I, II) and $\sigma_k{=}1$ for antisymmetric branches (Types~III, IV), yielding $r_k=\lceil N_k/2\rceil$ (Types~I, II, IV), $r_k= N_k/2+1$ (Type~III). For a general FIR branch, $r_k = 0$, thus $\mathbf{C}_k=\mathbf{0}_{0\times(N_k+1)}$ is the empty matrix~\cite[Ch.~2.1, p.~83]{bernstein2009matrix}, which contributes no rows to $\mathbf{C}$ but preserves the column partitioning, leaving $\mathbf{h}_k$ unconstrained.\footnote{Alternatively, a branch can be excluded from the optimization, e.g., the pure delay branch when Types I and III filters are present, and fixed a priori \cite{Johansson2003}, and its contribution subtracted from $\mathbf{b}$ in \eqref{eq:b}.}
The global constraint matrix $\mathbf C {\in} \mathbb{R}^{(P{-}R){\times} P}$ [with $R$ as in \eqref{eq:free}] is obtained as the block--diagonal concatenation of the per--branch matrices
\begin{align}
      \mathbf{C}&=\operatorname{blkdiag}(\mathbf{C}_{0},\dots,\mathbf{C}_{L}),
\end{align}
with the feasible coefficient vectors given by the nullspace 
\begin{align}
      \hspace{-4 pt}\mathcal N(\mathbf C) {=} \{\boldsymbol{\theta}{\in}\mathbb{R}^P\hspace{-5 pt}: \mathbf{C}\boldsymbol{\theta} {=} \mathbf 0\}
      {=} \operatorname{range}(\mathbf N)
      {\eqb} \{\mathbf N\boldsymbol{\beta}\hspace{-2 pt} : \boldsymbol{\beta}{\in}\mathbb{R}^R\},
      \label{eq:nullspace}
\end{align}
where $\mathbf N=[\mathbf n_1\ \cdots\ \mathbf n_R]\in\mathbb{R}^{P\times R}$ and $\mathbf C\mathbf n_r{=}\mathbf 0$ for $r=1,\dots,R$, with
\begin{align}
      R = \sum_{k=0}^{L} (N_k + 1 - r_k) = P - \sum_{k=0}^{L} r_k,
      \label{eq:free}
\end{align}
being the number of free design parameters after imposing the per-branch constraints, i.e., the dimension of $\mathcal N(\mathbf C)$.
Equality (b) in \eqref{eq:nullspace} [i.e., the representation $\mathcal N(\mathbf C) {=} \{\mathbf N\boldsymbol{\beta}: \boldsymbol{\beta}{\in}\mathbb{R}^R\}$] holds because $\mathbf N$ spans the nullspace of $\mathbf C$, so every $\boldsymbol{\theta}$ satisfying $\mathbf C\boldsymbol{\theta}{=}\mathbf0$ can be written as $\boldsymbol{\theta}{=}\mathbf N\boldsymbol{\beta}$ for some $\boldsymbol{\beta}{\in}\mathbb{R}^R$.\footnote{A basis $\mathbf{N}$ can be computed by solving $\mathbf C\mathbf N{=} \mathbf 0$, e.g., via the singular value decomposition (SVD)~\cite[Thm.~2.6.3, p.~150]{horn2012matrix} where $\mathbf{C} = \mathbf{U}\boldsymbol{\Sigma}\mathbf{V}^{\!\top}$ with $\boldsymbol{\Sigma} = \operatorname{diag}(\sigma_1,\dots,\sigma_{\operatorname{rank}(\mathbf{C})},0,\dots,0)$, then the last $R$ columns of $\mathbf{V}$ span $\mathcal{N}(\mathbf{C})$, alternatively via QR factorization with column pivoting of $\mathbf C$~\cite[Sec.~II.3, p.~143]{strang2019linear} or analytically by eq.~\eqref{eq:analytical_nullspace} in Appendix~\ref{app:explicit_nullspace}.}  
\begin{remark}[Branch Compatibility]
  For each branch, the nullspace basis is centered at $D$ to ensure consistent group delay; branches of different orders may be used provided this centering is applied. Types~I/III (even $N_k$, integer $D$) and Types~II/IV (odd $N_k$, half-integer $D$) cannot be mixed. General FIR branches ($r_k{=}0$) carry no such restriction.
\end{remark}
\begin{remark}[Existence of Nullspace]
      $\mathcal{N}(\mathbf{C})$ admits a real basis $\mathbf{N}{\in}\mathbb{R}^{P\times R}$ with $\dim\mathcal{N}(\mathbf{C}){=}R{>}0$; see Appendix~\ref{app:rank_nullity}. 
\end{remark}
Substituting the nullspace parameterization $\boldsymbol{\theta}=\mathbf N\boldsymbol{\beta}$, where $\boldsymbol{\beta}\in\mathbb{R}^R$ is the vector of $R$ free design parameters, in \eqref{eq:unconstrained}, the constrained least--squares problem becomes
\begin{align}
\boldsymbol{\theta}^\star
&= \arg\min_{\boldsymbol{\theta}:\mathbf{C}\boldsymbol{\theta}=\mathbf 0}
      \|\mathbf{A}\boldsymbol{\theta}-\mathbf{b}\|_2^2 \nonumber \\
&= \arg\min_{\boldsymbol{\theta}\in\mathcal{N}(\mathbf{C})}
      \|\mathbf{A}\boldsymbol{\theta}-\mathbf{b}\|_2^2 \nonumber \\
&= \arg\min_{\boldsymbol{\beta}}
      \|\mathbf{A}(\mathbf{N}\boldsymbol{\beta})-\mathbf{b}\|_2^2 .
\end{align}
For generally complex-valued $\mathbf{A}$ and $\mathbf{b}$, since $\boldsymbol{\theta}\in\mathbb{R}^P$ and $\mathbf{N}\in\mathbb{R}^{P\times R}$, we require $\boldsymbol{\beta}\in\mathbb{R}^R$. Separating real and imaginary parts of the complex residual gives the equivalent real least--squares problem

\begin{align}
\boldsymbol{\theta}^\star
&= \arg\min_{\boldsymbol{\beta}\in\mathbb{R}^R}
      \|\mathbf A_{\mathrm R}\mathbf N\boldsymbol{\beta}-\mathbf b_{\mathrm R}\|_2^2,
\end{align}
where
\begin{align}
\mathbf A_{\mathrm R} \triangleq
      \begin{bmatrix}
      \Re\{\mathbf A\}\\
      \Im\{\mathbf A\}
      \end{bmatrix},
\qquad
\mathbf b_{\mathrm R} \triangleq
      \begin{bmatrix}
      \Re\{\mathbf b\}\\
      \Im\{\mathbf b\}
      \end{bmatrix}.
\end{align}
The resulting closed--form solution is therefore given by
\begin{align}
\boldsymbol{\theta}^\star
= \mathbf N \boldsymbol{\beta}^\star,\
\boldsymbol{\beta}^\star=
      \bigl[(\mathbf{A_{\mathrm R}}\mathbf{N})^{\top}(\mathbf{A_{\mathrm R}}\mathbf{N})\bigr]^{-1}
      (\mathbf{A_{\mathrm R}}\mathbf{N})^{\top}\mathbf{b_{\mathrm R}},
      \label{eq:proposed}
\end{align}
assuming that $\mathbf A_{\mathrm R}\mathbf N$ has full column rank. This condition is satisfied for any nondegenerate frequency--delay grid with $M{\ge}R$ (by the same argument as Footnote~\ref{foo:grid}, with $P$ replaced by $R$). This is less restrictive than the unconstrained condition $M {\geq} P$ for \eqref{eq:unconstrained}, as $P {\geq} R$. Therefore, imposing linear-phase symmetry reduces both the multiplication count and the minimum grid density required for a unique solution.

\begin{remark} [Uniqueness]
      Although the nullspace basis $\mathbf N$ is not unique, the least--squares minimizer $\boldsymbol{\theta}^\star$ in \eqref{eq:proposed} is unique and independent of the chosen basis. A proof is given in Appendix~\ref{app:uniqueness}.
\end{remark}
\begin{remark}[Optimal Basis]
Although Appendix~\ref{app:uniqueness} guarantees that $\boldsymbol{\theta}^\star$ is independent of the chosen basis, the sensitivity of $\boldsymbol{\beta}^\star$ to perturbations in $\mathbf{A}_{\mathrm{R}}\mathbf{N}$ and $\mathbf{b}_{\mathrm{R}}$ is governed by \cite[eq.~(1.4.28), p.~31]{bjorck1996numerical}
\begin{align}
    \kappa_{\mathrm{LS}} = \kappa(\mathbf{A}_{\mathrm{R}}\mathbf{N})
    \left(1 + \kappa(\mathbf{A}_{\mathrm{R}}\mathbf{N})
    \frac{\|\mathbf{r}\|_2}{\|\mathbf{A}_{\mathrm{R}}\mathbf{N}\|_2\|\boldsymbol{\beta}^\star\|_2}
    \right),
\end{align}
where $\kappa(\cdot){=}\sigma_{\max}(\cdot)/\sigma_{\min}(\cdot)$ is the condition number and $\mathbf{r} = \mathbf{b}_{\mathrm{R}} - \mathbf{A}_{\mathrm{R}}\mathbf{N}\boldsymbol{\beta}^\star$ is the residual. Since $\kappa_{\mathrm{LS}}$ grows with $\kappa(\mathbf{A}_{\mathrm{R}}\mathbf{N})$, minimizing $\kappa(\mathbf{A}_{\mathrm{R}}\mathbf{N})$ over all valid bases $\mathbf{N}$ --which preserves the rank of $\mathbf{A}_{\mathrm{R}}\mathbf{N}$ (Appendix~\ref{app:rank_invariance}) but not its singular values-- reduces the sensitivity of \eqref{eq:proposed} to numerical errors.
Let $\mathbf{A}_{\mathrm{R}}\mathbf{N} = \mathbf{U}\boldsymbol{\Sigma}\mathbf{V}^\top$ be the SVD. Choosing $\mathbf{N}_{\mathrm{opt}} = \mathbf{N}\mathbf{V}\boldsymbol{\Sigma}^{+}$ yields $(\mathbf{A}_{\mathrm{R}}\mathbf{N}_{\mathrm{opt}})^\top (\mathbf{A}_{\mathrm{R}}\mathbf{N}_{\mathrm{opt}}) = \mathbf{I}$, achieving $\kappa = 1$, the global minimum, and \eqref{eq:proposed} reduces to
\begin{align}
      \boldsymbol{\theta}^\star = \mathbf{N}_{\mathrm{opt}}\boldsymbol{\beta}^\star, \quad
      \boldsymbol{\beta}^\star = \mathbf{V}\boldsymbol{\Sigma}^{+}\mathbf{U}^\top
    \mathbf{b}_{\mathrm{R}}, 
\end{align}
where $\boldsymbol{\Sigma}^+$ is the Moore--Penrose pseudoinverse. Equivalently, $\mathbf{N}_{\mathrm{opt}} = \mathbf{N}\mathbf{L}^{-1}$ where $\mathbf{L}$ is the Cholesky factor of $(\mathbf{A}_{\mathrm{R}}\mathbf{N})^\top(\mathbf{A}_{\mathrm{R}}\mathbf{N})$, which is cheaper to compute but requires the Gram matrix to be positive definite.
\end{remark}

\begin{remark}[Weighted Least Squares]
The formulation extends straightforwardly to a weighted least--squares (WLS) criterion by replacing $\mathbf{A}_{\mathrm{R}}$ and $\mathbf{b}_{\mathrm{R}}$ with $\mathbf{W}^{1/2}\mathbf{A}_{\mathrm{R}}$ and $\mathbf{W}^{1/2}\mathbf{b}_{\mathrm{R}}$, respectively, where $\mathbf{W}$ is a diagonal matrix of frequency--delay weights; all structural properties of the solution are preserved.
\end{remark}

\begin{algorithm}[tbp]
\caption{Branch-order selection}
\label{alg:design}
\begin{algorithmic}[1]
\REQUIRE $L$, $D$, $\varepsilon$, Linear phase types $\{\mathcal{T}_k^0\}_{k=1}^{L}$
\STATE Find minimum $N$ s.t.\ $N_k{=}N$, $\mathcal{T}_k{=}\mathcal{T}_k^0$, $\|e\|_2 \leq \varepsilon$
\STATE For $k{=}L,\dots,1$: decrease $N_k{\leftarrow}N_k{-}2$ while $\|e\|_2{\leq}\varepsilon$
\STATE $\forall k$: if $N_k/2 > D$, set $\mathcal{T}_k{\leftarrow}\mathrm{G}$, $N_k{\leftarrow}2D$
\STATE While $\|e\|_2 > \varepsilon$: $N_k{\leftarrow}N_k{+}1$ for each G-branch
\STATE $k{=}L,\dots,1$: if G-branch and $\mathcal{T}_k^0 \neq G$: try $\mathcal{T}_k{\leftarrow}\mathcal{T}_k^0$, $N_k{\leftarrow}2D$; if $\|e\|_2{\leq}\varepsilon$ accept and apply Step~2, else revert
\STATE $k{=}L,\dots,1$: if G-branch $N_k{\leftarrow}N_k{-}1$ while $\|e\|_2{\leq}\varepsilon$
\RETURN $\boldsymbol{\theta}^\star$ via~\eqref{eq:proposed}
\end{algorithmic}
\end{algorithm}

\section{Numerical Results} \label{sec:results}
\example\label{ex:caseA} 
We consider the design of a fractional-delay filter achieving $\|e\|_2{\leq}{-}100$~dB over $\Omega{=}[-0.9\pi,\,0.9\pi]$ and $\mathcal{D}{=}[-0.5,\,0.5]$ subject to a group delay constraint $D {=} 20$, using $N_\omega{=}1000$ frequency samples and $N_d{=}500$ delay values. Two feasible designs are compared in Table~\ref{tab:comparison} as detailed below; the all linear-phase configuration is included for reference only, as both the proposed framework restricted to linear-phase branches and the method of~\cite{Deng2006} yield identical results ($D{=}29$, $\|e\|_2{=}{-}100.56$~dB), confirming the equivalence stated in Section~\ref{sec:intro}, but do not satisfy the group delay $D{=}20$. In all designs, the branch types, $N_k$, and free parameters $R$ characterize the $L{=}6$ branches $k{=}1,\dots,6$, while the $k{=}0$ branch becomes a pure delay. In addition, Table~\ref{tab:complexity} summarizes the implementation complexity of the subfilters, i.e., $H_k(z)$ in Fig.~\ref{flo:farrow}, for both feasible designs using \mbox{multiply-add} operation counts with coefficient symmetry reduction, and delay and addition sharing between the branches.

\textbf{All general FIR branches.} The baseline applies our framework through Algorithm~\ref{alg:design} with $r_k{=}0$ (forcing no linear-phase constraints), which reduces to the standard unconstrained LS solution~\eqref{eq:unconstrained}. The algorithm produces per-branch orders $N_k{=}[62, 61, 62, 45, 39, 29]$ (with $L=6$), achieving $\Vert e\Vert_2 = -100.00$ dB at the cost of $R=304$ free parameters. This design satisfies the group delay constraint serving as a performance upper bound.\footnote{Optimizing the $k{=}0$ branch as a free general FIR filter as well (seven branches total) yields the same $\|e\|_2{=}{-}100.00$~dB at higher complexity; fixing the $k{=}0$ branch as a pure delay does not impose any restriction.}

\begin{figure}[tp]
	\centering
	\includegraphics[scale=0.97]{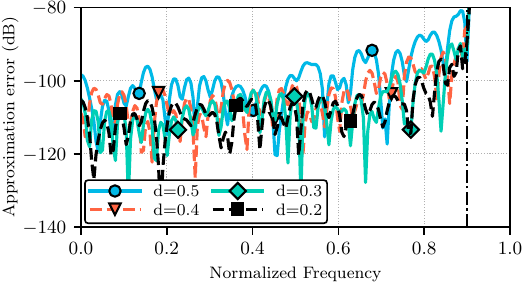}
	\caption{\textbf{\textit{Example \ref{ex:caseA}: Mixed general and linear-phase FIR branches.} }Approximation error $\vert H(e^{j\omega},d){-}H_{des}(e^{j \omega}, d) \vert $ for different fractional \mbox{delay values.}}
	\label{flo:comparisson}
\end{figure}

\textbf{Mixed general and linear-phase FIR branches.} The proposed framework via Algorithm~\ref{alg:design} yields mixed branch types $[\mathrm{G,I,G,I,III,I}]$, where $\mathrm{G}$ stands for general FIR, with per-branch orders $N_k{=}[63, 40, 56, 30, 36, 10]$, satisfying the targeted group delay $D{=}20$ and achieving $\|e\|_2{=}{-}100.01$~dB with $R{=}182$ free parameters, a $40\%$ reduction relative to the general FIR baseline. The two G-type branches ($k{=}1,3$) carry no linear-phase group-delay constraint; orders $N_k{=}63$ and $56$ are therefore admissible within $D{=}20$. Fig.~\ref{flo:comparisson} shows the approximation error of this design for different fractional delay values $d\in \{0.2,\,0.3,\,0.4,\,0.5\}$ and $\omega \in [0, 0.9\pi]$.

\newcommand{\bwidth}{1.0em}  
\newcommand{\branchsep}{0.3pt} 
\newcommand{\blocksepa}{12pt}  
\newcommand{\blocksepb}{8pt}  
\newcommand{\blocksepd}{-5pt}  
\newcolumntype{B}{>{\centering\arraybackslash}p{\bwidth}}
\begin{table}[tp]
\centering
\small
\caption{\textbf{\textit{Example \ref{ex:caseA}:} }Design comparison: $\Omega{=}[-0.9\pi,\,0.9\pi]$, $\mathcal{D}{=}[-0.5,0.5]$}
\label{tab:comparison}
\begin{tabular}{l @{\hspace{\blocksepa}} *{5}{B @{\hspace{\branchsep}}} B @{\hspace{\blocksepa}} l@{\hspace{\blocksepb}} c @{\hspace{\blocksepb}} l}
\toprule
Subfilters Design &  \multicolumn{6}{@{\hspace{\blocksepd}}c}{FIR Branch types\textsuperscript{$\dagger$}} & $D$ & $R$ & $\|e\|_2$ (dB) \\
\midrule
General            & G   & G & G   & G & G   & G & 20                            & 304          & $-100.00$ \\
Linear phase~\cite{Deng2006} & III & I & III & I & III & I & 29\textsuperscript{$\ddagger$} & 177          & $-100.56$ \\
Mixed                & G   & I & G   & I & III & I & \textbf{20}                   & \textbf{182} & $-100.01$ \\
\bottomrule
\noalign{\vspace{2pt}}
\multicolumn{10}{l}{\textsuperscript{$\dagger$}G\,=\,general FIR; I\,=\,Type-I; III\,=\,Type-III.}\\
\multicolumn{10}{l}{\textsuperscript{$\ddagger$}Does not satisfy the targeted $D{=}20$.}
\end{tabular}
\end{table}
\newcommand{\blocksepc}{14pt}  
\begin{table}[tp]
\centering
\small
\caption{\textbf{\textit{Example \ref{ex:caseA}:} }Implementation complexity}
\label{tab:complexity}
\begin{tabular}{l@{\hspace{\blocksepc}} r @{\hspace{\blocksepc}} r@{\hspace{\blocksepc}} r<{\hspace{5pt}}}
\toprule
Subfilters Design & \multicolumn{1}{r}{\hspace{6pt}Delays} & \multicolumn{1}{c}{\hspace{14pt}Mult\textsuperscript{$\star$}} & \multicolumn{1}{r}{\hspace{10pt}Add\textsuperscript{$\star$}} \\
\midrule
General & 62 & 304 & 298 \\
Mixed & \textbf{63} & \textbf{182} & \textbf{196} \\
\midrule
Saving & (-1.6\%)~~-1 & (40\%)~122 & (34\%)~102 \\
\bottomrule 
\noalign{\vspace{2pt}}
\multicolumn{4}{@{}l@{}}{\parbox{8.3cm}{\textsuperscript{$\star$}The six multiply-add operations for the Horner
polynomial-in-$d$ evaluation are common for all cases and therefore excluded.}}\\
\end{tabular} 
\end{table}

\example\label{ex:caseB} We consider the same frequency-delay grid as in \textit{Example~\ref{ex:caseA}},~but with an approximation error of \mbox{$\|e\|_2{\leq}-80$~dB} and a tighter group delay constraint $D {=} 14.5$. To meet the specification with general FIR filters, using $L{=}6$, the minimum per-branch orders are $N_k{=}[45, 45, 43, 45, 32, 28, 16]$ at the cost of $R{=}261$ free parameters. The proposed framework yields the mixed branch configuration $[\mathrm{G, G, G, IV, G, IV, II}]$ with per-branch orders $N_k{=}[47, 46, 46, 29, 34, 13, 3]$, and $R{=}201$ free parameters (a $23\%$ reduction relative to the general FIR baseline) while both satisfying the targeted group delay at $D{=}14.5$. The all linear-phase configuration requires a group delay $D{=}21.5$ to meet the same specifications, therefore, also failing to satisfy the delay constraint.  

\section{Conclusion} \label{sec:conclusion}
This letter presented a closed-form solution via a unified nullspace-based least-squares design for Farrow structures that accommodates mixed linear-phase and general FIR subfilters, as required when a group-delay constraint is imposed. Linear-phase branches of order $N_k$ impose a group delay of $N_k/2$; meeting a tight group-delay constraint at the target accuracy therefore requires mixing linear-phase branches with general FIR branches, for which no linear-phase group-delay restriction applies when $D {<} N_k/2$. The proposed solution coincides with~\cite{Tarczynski1997} and~\cite{Deng2006} in the all-general-FIR and all-linear-phase cases, respectively. Examples showed that the proposed mixed linear-phase and general-FIR framework achieves the same approximation error $\big(\|e\|_2$ sense$\big)$ as the all-general-FIR baseline while also meeting the targeted group delay, but with a significant reduction in the number of free parameters.

\appendices
\section{}\label{app:rank_nullity}
\noindent \textbf{Rank Deficiency of \texorpdfstring{$\mathbf{C}$}{C} and Existence of a Nullspace Basis:} For each branch $k=0,\dots,L$, $\mathbf{C}_k\in\mathbb{R}^{r_k\times(N_k+1)}$ encodes $r_k$ symmetry constraints ($r_k{=}0$ for general FIR branches), so there exists $\mathbf{N}_k\in\mathbb{R}^{(N_k+1)\times(N_k+1-r_k)}$ with $\mathbf{C}_k\mathbf{N}_k{=}\mathbf{0}$ and $\operatorname{rank}(\mathbf{N}_k){=}N_k{+}1{-}r_k$. Setting $\mathbf{N}{=}\operatorname{blkdiag}(\mathbf{N}_0,\dots,\mathbf{N}_L)$ and $\mathbf{C}{=}\operatorname{blkdiag}(\mathbf{C}_0,\dots,\mathbf{C}_L)$ gives $\mathbf{C}\mathbf{N}{=}\mathbf{0}$ and $\operatorname{rank}(\mathbf{N})=R=\sum_k(N_k{+}1{-}r_k)=P{-}\sum_k r_k$. Since $\operatorname{rank}(\mathbf{C}){=}P{-}R$, by the rank--nullity theorem \cite[Thm.~2.3, p.~70]{friedberg2014linear}, $\dim\mathcal{N}(\mathbf{C}){=}R$. Since the $R$ columns of $\mathbf{N}$ lie in $\mathcal{N}(\mathbf{C})$, therefore, $\operatorname{range}(\mathbf{N}){=}\mathcal{N}(\mathbf{C})$, confirming $\mathbf{N}$ is a valid nullspace basis. When all $r_k{=}0$, $R{=}P$ and~\eqref{eq:proposed} reduces to~\eqref{eq:unconstrained}.

\vspace{-5pt}
\section{} \label{app:uniqueness}
\noindent \textbf{Uniqueness:} Let $\mathbf{N}_{\mathrm{I}}$ and $\mathbf{N}_{\mathrm{II}}$ be two real bases of $\mathcal{N}(\mathbf{C})$, there exists an invertible matrix $\mathbf{T}\in\mathbb{R}^{R\times R}$ such that $\mathbf{N}_{\mathrm{II}} = \mathbf{N}_{\mathrm{I}}\mathbf{T}$. Substituting $\mathbf{N}_{\mathrm{II}}$ into \eqref{eq:proposed} gives
\begin{align*}
\boldsymbol{\theta}^\star_{\mathrm{II}} &= \mathbf{N}_{\mathrm{II}}\boldsymbol{\beta}_{\mathrm{II}}^\star \\
      &= \mathbf{N}_{\mathrm{I}}\mathbf{T}\bigl(\mathbf{T}^{\!\top}\mathbf{N}_{\mathrm{I}}^{\!\top}\mathbf{A}_{\mathrm{R}}^{\!\top}
      \mathbf{A}_{\mathrm{R}}\mathbf{N}_{\mathrm{I}}\mathbf{T}\bigr)^{-1}
      \mathbf{T}^{\!\top}\mathbf{N}_{\mathrm{I}}^{\!\top}\mathbf{A}_{\mathrm{R}}^{\!\top}\mathbf{b}_{\mathrm{R}} \\
      &=\mathbf{N}_{\mathrm{I}}\mathbf{T}\big[\mathbf{T}^{-1}\bigl(\mathbf{N}_{\mathrm{I}}^{\!\top}\mathbf{A}_{\mathrm R}^{\!\top}\mathbf{A}_{\mathrm R}\mathbf{N}_{\mathrm{I}}\bigr)^{-1}(\mathbf{T}^{\!\top})^{-1}\big]\mathbf{T}^{\!\top}\mathbf{N}_{\mathrm{I}}^{\!\top}\mathbf{A}_{\mathrm{R}}^{\!\top}\mathbf{b}_{\mathrm{R}}, \\
      &\eqa \mathbf{N}_{\mathrm{I}}\boldsymbol{\beta}_{\mathrm{I}}^\star = \boldsymbol{\theta}^\star_{\mathrm{I}}, \tag{B1}      
\end{align*}
since $\mathbf{T}$ cancels out, $\boldsymbol{\theta}^\star$ is unique, and depends only on the subspace $\mathcal{N}(\mathbf{C})$, not on the particular basis used to represent it. Here, step (a) followed from applying \eqref{eq:proposed}.

\section{}\label{app:rank_invariance}
\noindent \textbf{Rank Invariance Under Basis Change:} By Appendix~\ref{app:uniqueness}, $\mathbf{N}_{\mathrm{II}} = \mathbf{N}_{\mathrm{I}}\mathbf{T}$ for some invertible $\mathbf{T}\in\mathbb{R}^{R\times R}$. Applying 
$\mathrm{rank}(\mathbf{AB}) \leq \min\{\mathrm{rank}(\mathbf{A}), \mathrm{rank}(\mathbf{B})\}$
\begin{align*}
      \mathrm{rank}(\mathbf{A}_{\mathrm{R}}\mathbf{N}_{\mathrm{II}}) &= \mathrm{rank}(\mathbf{A}_{\mathrm{R}}\mathbf{N}_{\mathrm{I}}\mathbf{T}) \tag{C1}\\ 
      &\leq \min \{\mathrm{rank}(\mathbf{A}_{\mathrm{R}}\mathbf{N}_{\mathrm{I}}), \,\, \mathrm{rank}(\mathbf{T})\} \\
      &\leq \mathrm{rank}(\mathbf{A}_{\mathrm{R}}\mathbf{N}_{\mathrm{I}}) \tag{C2}\\
      &= \mathrm{rank}(\mathbf{A}_{\mathrm{R}}\mathbf{N}_{\mathrm{II}}\mathbf{T}^{-1}), \quad \mathbf{N}_{\mathrm{I}}= \mathbf{N}_{\mathrm{II}}\mathbf{T}^{-1}\\
      & \leq \min \{\mathrm{rank}(\mathbf{A}_{\mathrm{R}}\mathbf{N}_{\mathrm{II}}), \,\, \mathrm{rank}(\mathbf{T}^{-1})\} \\
      & \leq \mathrm{rank}(\mathbf{A}_{\mathrm{R}}\mathbf{N}_{\mathrm{II}}), \tag{C3}
\end{align*}
which, by (C1)--(C3) implies $\mathrm{rank}(\mathbf{A}_{\mathrm{R}}\mathbf{N}_{\mathrm{I}}) = \mathrm{rank}(\mathbf{A}_{\mathrm{R}}\mathbf{N}_{\mathrm{II}})$, i.e., the rank of $\mathbf{A}_{\mathrm{R}}\mathbf{N}$ is invariant under change of basis.

\section{} \label{app:explicit_nullspace}
\noindent \textbf{Explicit Nullspace Construction:} Define the per-branch basis $\mathbf{S}_k\in\mathbb{R}^{(N_k+1)\times(N_k+1-r_k)}$  via its columns [cf.~\eqref{eq:Ck_entries}, with the sign of~$\sigma_k$ changed]
\begin{equation}
  \mathbf{s}_m^{k}[n] = \delta(n{-}m) - \sigma_k\,\delta\!\bigl(n{-}(N_k{-}m)\bigr), \tag{D1}
  \label{eq:nullspace_basis_vec}
\end{equation}
with $m{=}0,\dots,\lceil N_k/2\rceil{-}1$ and the center column $\delta(n{-}N_k/2)$ appended for Type~I (even $N_k$, ${-}\sigma_k{=}1$); for general FIR branches, $\mathbf{S}_k{=}\mathbf{I}_{N_k+1}$. The global nullspace basis is
\begin{equation}
  \mathbf{N} = \operatorname{blkdiag}(\mathbf{S}_{0},\dots,\mathbf{S}_{L})
  \in\mathbb{R}^{P\times R}, \tag{D2}
  \label{eq:analytical_nullspace}
\end{equation}
which satisfies $\mathbf{C}\mathbf{N}=\mathbf{0}$ by construction.

\begin{small} 
	\bibliographystyle{IEEEtran}
   \bibliography{IEEEabrv, references/references_main} 

\begin{thebibliography}{10}
\providecommand{\url}[1]{#1}
\csname url@samestyle\endcsname
\providecommand{\newblock}{\relax}
\providecommand{\bibinfo}[2]{#2}
\providecommand{\BIBentrySTDinterwordspacing}{\spaceskip=0pt\relax}
\providecommand{\BIBentryALTinterwordstretchfactor}{4}
\providecommand{\BIBentryALTinterwordspacing}{\spaceskip=\fontdimen2\font plus
\BIBentryALTinterwordstretchfactor\fontdimen3\font minus \fontdimen4\font\relax}
\providecommand{\BIBforeignlanguage}[2]{{%
\expandafter\ifx\csname l@#1\endcsname\relax
\typeout{** WARNING: IEEEtran.bst: No hyphenation pattern has been}%
\typeout{** loaded for the language `#1'. Using the pattern for}%
\typeout{** the default language instead.}%
\else
\language=\csname l@#1\endcsname
\fi
#2}}
\providecommand{\BIBdecl}{\relax}
\BIBdecl

\bibitem{Chen2024}
Y.-M. Chen and C.-C. Chen, ``Design of {F}arrow structured variable fractional delay filter for time-varying {LEO} communication channel emulator with {SRRC} communication waveforms,'' \emph{IEEE Access}, vol.~12, pp. 122\,229--122\,238, Aug. 2024.

\bibitem{Liu2015}
Z.~Liu, Y.~L. Guan, and H.-H. Chen, ``Fractional-delay-resilient receiver design for interference-free {MC-CDMA} communications based on complete complementary codes,'' \emph{IEEE Trans. Wirel. Commun.}, vol.~14, no.~3, pp. 1226--1236, Oct. 2015.

\bibitem{deiokshak2025}
D.~Rodriguez-Linares, O.~Moryakova, and H.~Johansson, ``Joint sampling frequency offset estimation and compensation based on the {F}arrow structure,'' in \emph{Proc. 25th Int. Conf. Digit. Signal Process. (DSP)}, Costa Navarino, Messinia, Greece, Jun. 2025, pp. 1--5.

\bibitem{Shi_2023}
C.~Shi, X.~Xie, X.~Zhang, and L.~Yu, ``Calibration of timing mismatch for {TIADC} based on error table and fractional delay filter,'' \emph{J. Phys.: Conf. Ser.}, vol. 2525, no.~1, p. 012001, Jun. 2023.

\bibitem{Canese2023}
L.~Canese, G.~C. Cardarilli, L.~Di~Nunzio, R.~Fazzolari, D.~Giardino, M.~Re, and S.~Spanò, ``Efficient digital implementation of a multirate-based variable fractional delay filter for wideband beamforming,'' \emph{IEEE Trans. Circuits Syst. II: Express Briefs}, vol.~70, no.~6, pp. 2231--2235, Jan. 2023.

\bibitem{Farrow1988}
C.~Farrow, ``A continuously variable digital delay element,'' in \emph{IEEE Int. Symp. Circuits Syst. (ISCAS)}, vol.~3, Espoo, Finland, Aug. 1988, pp. 2641--2645.

\bibitem{Hunter2009}
M.~T. Hunter and W.~B. Mikhael, ``A novel {F}arrow structure with reduced complexity,'' in \emph{Proc. 52nd IEEE Int. Midwest Symp. Circuits Syst. (MWSCAS)}, Cancun, Mexico, Sept. 2009, pp. 581--585.

\bibitem{Eghbali2013}
A.~Eghbali, H.~Johansson, and T.~Saramäki, ``A method for the design of {F}arrow-structure based variable fractional-delay {FIR} filters,'' \emph{Signal Process.}, vol.~93, no.~5, pp. 1341--1348, May 2013.

\bibitem{Johansson2003}
H.~Johansson and P.~L\"{o}wenborg, ``On the design of adjustable fractional delay {FIR} filters,'' \emph{IEEE Trans. Circuits Syst. II: Express Briefs}, vol.~50, no.~4, pp. 164--169, Apr. 2003.

\bibitem{Laakso1996}
T.~Laakso, V.~V\"{a}lim\"{a}ki, M.~Karjalainen, and U.~Laine, ``Splitting the unit delay.'' \emph{IEEE Signal Process. Mag.}, vol.~13, no.~1, pp. 30--60, Jan. 1996.

\bibitem{Haolin2017}
H.~Li, G.~Torfs, T.~Kazaz, J.~Bauwelinck, and P.~Demeester, ``{F}arrow structured variable fractional delay lagrange filters with improved midpoint response,'' in \emph{Proc. 40th Int. Conf. Telecommun. Signal Process. (TSP)}, Barcelona, Spain, Jul. 2017, pp. 506--509.

\bibitem{Haolin2019}
H.~Li, J.~Van~Kerrebrouck, J.~Bauwelinck, P.~Demeester, and G.~Torfs, ``Maximally flat and least-square co-design of variable fractional delay filters for wideband software-defined radio,'' \emph{J. Circuits Syst. Comput.}, vol.~28, no.~01, pp. 1\,950\,006:1--1\,950\,006:20, Oct. 2018.

\bibitem{Samadi2004}
S.~Samadi, M.~Ahmad, and M.~Swamy, ``Results on maximally flat fractional-delay systems,'' \emph{IEEE Trans. Circuits Syst. I, Reg. Papers}, vol.~51, no.~11, pp. 2271--2286, Nov. 2004.

\bibitem{Srivatsan2020}
K.~Srivatsan and N.~Venkatesan, ``Farrow structure based {FIR} filter design using hybrid optimization,'' \emph{AEU - Int. J. Electron. Commun.}, vol. 114, pp. 153\,020:1--153\,020:14, Feb. 2020.

\bibitem{Tarczynski1997}
A.~Tarczynski, G.~Cain, E.~Hermanowicz, and M.~Rojewski, ``{WLS} design of variable frequency response {FIR} filters,'' in \emph{Proc. IEEE Int. Symp. Circuits Syst. (ISCAS)}, vol.~4, Jun. 1997, pp. 2244--2247 vol.4.

\bibitem{Deng2006}
T.-B. Deng and Y.~Lian, ``Weighted-least-squares design of variable fractional-delay {FIR} filters using coefficient symmetry,'' \emph{IEEE Trans. Signal Process.}, vol.~54, no.~8, pp. 3023--3038, Aug. 2006.

\bibitem{Eghbali2011}
A.~Eghbali and H.~Johansson, ``Complexity reduction in low-delay {F}arrow-structure-based variable fractional delay {FIR} filters utilizing linear-phase subfilters,'' in \emph{Proc. 20th Eur. Conf. Circuit Theory Des. (ECCTD)}, Link{\"o}ping, Sweden, Aug. 2011, pp. 21--24.

\bibitem{bernstein2009matrix}
D.~Bernstein, \emph{Matrix Mathematics: Theory, Facts, and Formulas}, 2nd~ed.\hskip 1em plus 0.5em minus 0.4em\relax Princeton, NJ, USA: Princeton University Press, 2009.

\bibitem{horn2012matrix}
R.~A. Horn and C.~R. Johnson, \emph{Matrix Analysis}, 2nd~ed.\hskip 1em plus 0.5em minus 0.4em\relax New York, NY, USA: Cambridge University Press, 2013.

\bibitem{strang2019linear}
G.~Strang, \emph{Linear Algebra and Learning from Data}.\hskip 1em plus 0.5em minus 0.4em\relax Wellesley, MA, USA: Wellesley-Cambridge Press, 2019.

\bibitem{bjorck1996numerical}
{\AA}.~Bj{\"o}rck, \emph{Numerical Methods for Least Squares Problems}.\hskip 1em plus 0.5em minus 0.4em\relax Philadelphia, PA, USA: Society for Industrial and Applied Mathematics (SIAM), 1996.

\bibitem{friedberg2014linear}
S.~H. Friedberg, A.~J. Insel, and L.~E. Spence, \emph{Linear Algebra}, 4th~ed.\hskip 1em plus 0.5em minus 0.4em\relax Boston, MA, USA: Pearson, 2014.

\end{thebibliography}
\end{small}

\end{document}